\documentclass[english,12pt]{article}
\usepackage[cp1251]{inputenc}
\usepackage{babel}
\usepackage{amsfonts, amsmath}

\newcommand{\be}{\begin{equation}} \newcommand{\ee}{\end{equation}}

\begin{document}
\title{Quantum Mechanics at Planck's scale and Density Matrix} \thispagestyle{empty}

\author{A.E.Shalyt-Margolin\hspace{1.5mm}\thanks
{Phone (+375) 172 883438; e-mail: a.shalyt@mail.ru; alexm@hep.by
}, J.G.Suarez
\thanks{Phone (+375) 172 883438; e-mail: suarez@hep.by, jsuarez@mail.tut.by}
}
\date{}
\maketitle
 \vspace{-25pt}
{\footnotesize\noindent  National Center of Particles and High
Energy Physics, Bogdanovich Str. 153, Minsk 220040, Belarus\\
{\ttfamily{\footnotesize
}}

\rm\normalsize \vspace{0.5cm}
\begin{abstract}
In this paper Quantum Mechanics with Fundamental Length is chosen
as Quantum Mechanics at Planck's scale. This is possible due to
the presence in the theory of General Uncertainty Relations. Here
Quantum Mechanics with Fundamental Length is obtained as a
deformation of Quantum Mechanics. The distinguishing feature of
the proposed approach in comparison with previous ones, lies on
the fact that here density matrix subjects to deformation whereas
so far commutators have been deformed. The density matrix obtained
by deformation of quantum-mechanical density one is named
throughout this paper density pro-matrix. Within our approach two
main features of Quantum Mechanics are conserved: the
probabilistic interpretation of the theory and the well-known
measuring procedure corresponding to that interpretation. The
proposed  approach allows to describe dynamics. In particular, the
explicit form of deformed Liouville's equation and  the deformed
Shr\"odinger's picture are given. Some implications of obtained
results are discussed. In particular, the problem of singularity,
the hypothesis of cosmic censorship, a possible improvement of the
definition of statistical entropy and the problem of information
loss in black holes are considered. It is shown that obtained
results allow to deduce in a simple and natural way the
Bekenstein-Hawking's formula for black hole entropy in
semiclassical approximation.
\end{abstract}

\section{Introduction}
In this paper Quantum Mechanics with Fundamental Length (QMFL) is
considered as Quantum Mechanics (QM) of the early Universe. The
main motivation for this choice  is the presence in the theory of
 General Uncertainty Relations (GUR) appropriated to describe the
 physical behavior of the early Universe
and unavoidable conducting to the concept of fundamental length.
Here QMFL is obtained as a deformation of QM, choosing the
quantity $\beta=l_{min}^{2}/x^{2}$ (where $x$ is the scale) as the
parameter of deformation of the theory. The main difference
between used
 approach and previous ones lies on the fact, that we propose to deform
density matrix  whereas so far commutators have been deformed.
Obtained in such a way density matrix (generalized density matrix)
is called here and throughout this paper density pro-matrix.
Within our approach two very important features of QM have been
conserved in QMFL. Namely,
 the probabilistic interpretation and the well-known measuring procedure, corresponding to this
interpretation. It was shown that in the paradigm of expanding
model of the Universe there are two different (unitary
non-equivalent) Quantum Mechanics: the first one named QMFL is
describing nature at Planck's scale or on the early Universe and
it is based on GUR. The second one named
 QM and representing passage to the limit from Planck's to low
energy scale is based on Heisenberg's Uncertainty Relations (UR).
Consequently, some well-known quantum mechanical concepts could
appear only in the low energy limit. Further, within the proposed
approach some dynamical aspects of QMFL are described. In
particular, a deformation of the Liouville's equation, the
Shr\"odinger's picture in QMFL as well as some implications of
obtained results are presented. Mentioned implications deal with
the problem of singularity,
 the hypothesis of cosmic censorship, a possible improvement of
definition of statistical
 entropy  and also with the problem of information
loss in black holes. The Bekenstein-Hawking's formula for black
hole entropy in semiclassical approximation
 is deduced within proposed approach in a simple and natural way.

\section{Fundamental Length and Density Matrix}

Using different approaches (String Theory \cite{r2}, Gravitation
\cite{r3}, Quantum Theory of black holes \cite{r4}  etc.) the
authors of numerous papers issued over the last 14-15 years have
pointed out that Heisenberg's Uncertainty Relations should be
modified. Specifically, a high energy correction has to appear
\begin{equation}\label{U2}
\triangle x\geq\frac{\hbar}{\triangle p}+\alpha
L_{p}^2\frac{\triangle p}{\hbar}.
\end{equation}
\noindent Here $L_{p}$ is the Planck's length:
$L_{p}=\sqrt\frac{G\hbar}{c^3}\simeq1,6\;10^{-35}\;m$ and
 $\alpha > 0$ is a constant. In \cite{r3} it was shown
that this constant may be chosen equal to 1. However, here we will use
$\alpha$ as an arbitrary constant without giving it any definite
value.
The  inequality (\ref{U2}) is quadratic in
$\triangle p$:
\begin{equation}\label{U3}
\alpha L_{p}^2({\triangle p})^2-\hbar \triangle x \triangle p+
\hbar^2 \leq0,
\end{equation}
from whence the fundamental length is
\begin{equation}\label{U4}
\triangle x_{min}=2\sqrt\alpha L_{p}.
\end{equation}
Since in what follows we proceed only from the existence of
fundamental length, it should be noted that this fact was
established apart from GUR as well. For instance, from an ideal
experiment associated with Gravitational Field and Quantum
Mechanics a lower bound on minimal length was obtained in
\cite{r6}, \cite{r7} and  improved in \cite{r8} without using GUR
to an estimate of the form $\sim L_{p}$. \noindent Let us to
consider equation (\ref{U4}) in some detail. Squaring both its
sides, we obtain
\begin{equation}\label{U5}
(\overline{\Delta\widehat{X}^{2}})\geq 4\alpha L_{p}^{2 },
\end{equation}
Or in terms of density matrix
\begin{equation}\label{U6}
Sp[(\rho \widehat{X}^2)-Sp^2(\rho \widehat{X}) ]\geq 4\alpha
L_{p}^{2 }=l^{2}_{min}>0,
\end{equation}
where $\widehat{X}$ is the coordinate operator. Expression
(\ref{U6}) gives the measuring rule used in QM. However, in the
case considered here, in comparison with QM, the right part of
(\ref{U6}) cannot be done arbitrarily near to zero since it is
limited by $l^{2}_{min}>0$, where due to GUR $l_{min} \sim L_{p}$.

 Apparently, this may be due to
the fact that QMFL with GUR (\ref{U2}) is unitary non-equivalent
to  QM with UR. Actually, in QM the left-hand side of (\ref{U6})
can be chosen arbitrary closed to zero, whereas in QMFL this is
impossible. But if two theories are unitary equivalent then, the
form of their spurs should be retained. Besides, a more important
aspect is contributing to unitary non-equivalence of these two
theories: QMFL contains three fundamental constants (independent
parameters) $G$, $c$ and $\hbar$, whereas QM contains only one:
$\hbar$. Within an inflationary model (see \cite{r9}), QM is the
low-energy limit of QMFL (QMFL turns to QM) for the expansion of
the Universe. In this case, the second term in the right-hand side
of (\ref{U2}) vanishes and GUR turn to UR. A natural way for
studying QMFL is to consider this theory as a deformation of QM,
turning to QM at the low energy limit (during the expansion of the
Universe after the Big Bang). We will consider precisely this
option. However differing from authors of papers \cite{r4},
\cite{r5} and others, we do not deform commutators, but density
matrix, leaving at the same time the fundamental
quantum-mechanical measuring rule (\ref{U6}) without changes. Here
the following question may be formulated: how should be deformed
density matrix conserving quantum-mechanical measuring rules in
order to obtain self-consistent measuring procedure in QMFL? For
answering to the question we will use the R-procedure. For
starting let us to consider R-procedure both at the Planck's
energy scale and at the low-energy one. At the Planck's scale $a
\approx il_{min}$ or $a \sim iL_{p}$, where $i$ is a small
quantity. Further $a$ tends to infinity and we obtain for density
matrix $$Sp[\rho a^{2}]-Sp[\rho a]Sp[\rho a] \simeq
l^{2}_{min}\;\; or\;\; Sp[\rho]-Sp^{2}[\rho] \simeq
l^{2}_{min}/a^{2}.$$

 Therefore:

 \begin{enumerate}
 \item When $a < \infty$, $Sp[\rho] =
Sp[\rho(a)]$ and
 $Sp[\rho]-Sp^{2}[\rho]>0$. Then, \newline $Sp[\rho]<1$
 that corresponds to the QMFL case.
\item When $a = \infty$, $Sp[\rho]$ does not depend on $a$ and
$Sp[\rho]-Sp^{2}[\rho]\rightarrow 0$. Then, $Sp[\rho]=1$ that
corresponds to the QM case.
\end{enumerate}
How should be points 1 and 2 interpreted? How does analysis
above-given agree to the main result from \cite{r29} \footnote
{"... there cannot be any physical state which is a position
eigenstate since a eigenstate would of course have zero
uncertainty in position"}? It is in full agreement. Indeed, when
state-vector reduction (R-procedure) takes place in QM then,
always an eigenstate (value) is chosen exactly. In other words,
the probability is equal to 1. As it was pointed out in the
above-mentioned point 1 the situation changes when we consider
QMFL: it is impossible to measure coordinates exactly since it
never will be absolutely reliable. We obtain in all cases a
probability less than 1 ($Sp[\rho]=p<1$). In other words, any
R-procedure in QMFL leads to an eigenvalue, but only with a
probability less than 1. This probability is as near to 1 as far
the difference between measuring scale $a$ and $l_{min}$ is
growing, or in other words, when the second term in (\ref{U2})
becomes insignificant and we turn to QM. Here there is not a
contradiction with \cite{r29}. In QMFL there are not exact
coordinate eigenstates (values) as well as there are not pure
states. In this paper we do not consider operator properties in
QMFL as it was done in \cite{r29} but density-matrix properties.

 The  properties of density matrix in
QMFL and QM have to be different. The only reasoning in this case
may be as follows: QMFL must differ from QM, but in such a way
that in the low-energy limit a density matrix in QMFL must
coincide with the density matrix in QM. That is to say, QMFL is a
deformation of QM and the parameter of deformation depends on the
measuring scale. This means that in QMFL $\rho=\rho(x)$, where $x$
is the scale, and for $x\rightarrow\infty$  $\rho(x) \rightarrow
\widehat{\rho}$, where $\widehat{\rho}$ is the density matrix in
QM.

Since on the Planck's scale $Sp[\rho]<1$, then for such scales
$\rho=\rho(x)$, where $x$ is the scale, is not a density matrix as
it is generally defined in QM. On Planck's scale we name $\rho(x)$
 "density pro-matrix". As follows from the above, the density
matrix $\widehat{\rho}$ appears in the limit
\begin{equation}\label{U12}
\lim\limits_{x\rightarrow\infty}\rho(x)\rightarrow\widehat{\rho},
\end{equation}
when GUR (\ref{U2}) turn to UR  and QMFL turns to QM.

Thus, on Planck's scale the density matrix is inadequate to obtain
all information about the mean values of operators. A "deformed"
density matrix (or pro-matrix) $\rho(x)$ with $Sp[\rho]<1$ has to
be introduced because a missing part of information $1-Sp[\rho]$
is encoded in the quantity $l^{2}_{min}/a^{2}$, whose specific
weight decreases as the scale $a$ expressed in  units of $l_{min}$
is going up.

\section{QMFL as a deformation of QM}
Here we describe QMFL as a deformation of QM using the
above-developed formalism of density pro-matrix. Within it density
pro-matrix should be understood as a deformed density matrix in
QMFL. As fundamental parameter of deformation we use the quantity
$\beta=l_{min}^{2 }/x^{2 }$, where $x$ is the scale.

\noindent {\bf Definition 1.}

\noindent Any system in QMFL is described by a density pro-matrix
of the form $\rho(\beta)=\sum_{i}\omega_{i}(\beta)|i><i|$, where
\begin{enumerate}
\item $0<\beta\leq1/4$;
\item Vectors $|i>$ form a full orthonormal system;
\item Coefficients $\omega_{i}(\beta)\geq 0$ and for all $i$  the
 limit $\lim\limits_{\beta\rightarrow
0}\omega_{i}(\beta)=\omega_{i}$ exists;
\item
$Sp[\rho(\beta)]=\sum_{i}\omega_{i}(\beta)<1$,
$\sum_{i}\omega_{i}=1$;
\item For every operator $B$ and any $\beta$ there is a
mean operator $B$ depending on  $\beta$:\\
$$<B>_{\beta}=\sum_{i}\omega_{i}(\beta)<i|B|i>.$$
\end{enumerate}
Finally, in order that our definition 1  agrees to the result of
section 2, the following condition must be fulfilled:
\begin{equation}\label{U13}
Sp[\rho(\beta)]-Sp^{2}[\rho(\beta)]\approx\beta.
\end{equation}
Hence we can find the value for
$Sp[\rho(\beta)]$ satisfying the condition of definition 1:
\begin{equation}\label{U14}
Sp[\rho(\beta)]\approx\frac{1}{2}+\sqrt{\frac{1}{4}-\beta}.
\end{equation}

According to point 5  $<1>_{\beta}=Sp[\rho(\beta)]$. Therefore,
for any scalar quantity $f$ we have $<f>_{\beta}=f
Sp[\rho(\beta)]$. In particular, the mean value
$<[x_{\mu},p_{\nu}]>_{\beta}$ is equal to
\begin{equation}\label{U15}
<[x_{\mu},p_{\nu}]>_{\beta}= i\hbar\delta_{\mu,\nu}
Sp[\rho(\beta)].
\end{equation}
We denote the limit $\lim\limits_{\beta\rightarrow
0}\rho(\beta)=\rho$ as the density matrix. Evidently, in the limit
$\beta\rightarrow 0$ we return to QM.

As follows from definition 1,
$<|j><j|>_{\beta}=\omega_{j}(\beta)$, from whence the completeness
condition by $\beta$ is
\\$<(\sum_{i}|i><i|)>_{\beta}=<1>_{\beta}=Sp[\rho(\beta)]$. The
norm of any vector $|\psi>$ assigned to  $\beta$ can be defined as

$$<\psi|\psi>_{\beta}=<\psi|(\sum_{i}|i><i|)_{\beta}|\psi>
=<\psi|(1)_{\beta}|\psi>=<\psi|\psi> Sp[\rho(\beta)],$$

 where
$<\psi|\psi>$ is the norm in QM, i.e. for $\beta\rightarrow 0$.
Similarly, the described theory may be interpreted using a
probabilistic approach, however requiring  replacement of $\rho$
by $\rho(\beta)$ in all formulae.

\renewcommand{\theenumi}{\Roman{enumi}}
\renewcommand{\labelenumi}{\theenumi.}
\renewcommand{\labelenumii}{\theenumii.}

It should be noted:

\begin{enumerate}
\item The above limit covers both Quantum
and Classical Mechanics. Indeed, since $\beta\sim L_{p}^{2 }/x^{2
}=G \hbar/c^3 x^{2 }$, we obtain:
\begin{enumerate}
\item $(\hbar \neq 0,x\rightarrow
\infty)\Rightarrow(\beta\rightarrow
0)$ for QM;
\item $(\hbar\rightarrow 0,x\rightarrow
\infty)\Rightarrow(\beta\rightarrow
0)$ for Classical Mechanics;
\end{enumerate}
\item As a matter of fact, the deformation parameter $\beta$
should assume the value $0<\beta\leq1$.  However, as seen from
(\ref{U14}), $Sp[\rho(\beta)]$ is well defined only for
$0<\beta\leq1/4$.That is if $x=il_{min}$ and $i\geq 2$ then, there
is not any problem. At the point where $x=l_{min}$ there is a
singularity related to complex values assumed by $Sp[\rho(\beta)]$
, i.e. to the impossibility of obtaining a diagonalized density
pro-matrix at this point over the field of real numbers. For this
reason definition 1 has no sense at the point $x=l_{min}$. We will
come back to the question appearing in this section when we will
discuss singularity and hypothesis of cosmic censorship in section
5.

\item We consider possible solutions for (\ref{U13}).
For instance, one of the solutions of (\ref{U13}), at least to the
first order in $\beta$, is $$\rho^{*}(\beta)=\sum_{i}\alpha_{i}
exp(-\beta)|i><i|,$$ where all $\alpha_{i}>0$ are independent of
$\beta$  and their sum is equal to 1. In this way
$Sp[\rho^{*}(\beta)]=exp(-\beta)$. Indeed, we can easily verify
that \begin{equation}\label{U15}
Sp[\rho^{*}(\beta)]-Sp^{2}[\rho^{*}(\beta)]=\beta+O(\beta^{2}).
\end{equation}
The exponential ansatz for $\rho^{*}(\beta)$ given here will be
taking into account in further sections. Note that in the momentum
representation $\beta=p^{2}/p^{2}_{pl}$, where $p_{pl}$ is the
Planck's momentum. When present in matrix elements, $exp(-\beta)$
can damp the contribution of great momenta in a perturbation
theory.
\item It is clear that within the proposed description the
states with a unit probability, i.e. pure states, can appear only
in the limit $\beta\rightarrow 0$, when all $\omega_{i}(\beta)$
except one are equal to zero or when they tend to zero at this
limit. In our treatment pure states are states, which can be
represented in the form $|\psi><\psi|$, where $<\psi|\psi>=1$.

\item We suppose that all definitions concerning a
density matrix can be carry over to the above-mentioned
deformation of Quantum Mechanics (QMFL)  changing the density
matrix $\rho$ by the density pro-matrix $\rho(\beta)$ and
subsequent passage to the low-energy limit $\beta\rightarrow 0$.
Specifically, for statistical entropy we have
\begin{equation}\label{U16}
S_{\beta}=-Sp[\rho(\beta)\ln(\rho(\beta))].
\end{equation}
The quantity of $S_{\beta}$ seems never to be equal to zero as
$\ln(\rho(\beta))\neq 0$ and hence $S_{\beta}$ may be equal
to zero at the limit $\beta\rightarrow 0$ only.
\end{enumerate}
Some Implications:
\begin{enumerate}
\item If we carry out a measurement at a pre-determined scale, it is
impossible to regard the density pro-matrix as a density matrix
with an accuracy better than the limit $\sim10^{-66+2n}$, where
$10^{-n}$ is the measuring scale. In the majority of known cases
this is sufficient to consider the density pro-matrix as a density
matrix. But at Planck's scale, where quantum gravitational effects
and Planck's energy levels cannot be neglected, the difference
between $\rho(\beta)$ and  $\rho$ should be taken into
consideration.

\item Proceeding from the above, on Planck's scale the
notion of Wave Function of the Universe (as introduced in
\cite{r10}) has no sense, and quantum gravitation effects in this
case should be described with the help of density pro-matrix
$\rho(\beta)$ only.
\item Since density pro-matrix $\rho(\beta)$ depends on the measuring
scale, evolution of the Universe within the inflationary model
paradigm \cite{r9} is not a unitary process, or otherwise the
probabilities $p_{i}=\omega_{i}(\beta)$  would be preserved.
\end{enumerate}

\section{Dynamical aspects of QMFL}
Let's suppose that in QMFL density pro-matrix has the form
$\rho[\beta(t),t]$ or in other words, it depends on two
parameters: time $t$ and parameter of deformation $\beta$, which
also depends on time ($\beta=\beta(t)$). Then, we have
\begin{equation}\label{U17}
\rho[\beta(t),t]=\sum\omega_{i}[\beta(t)]|i(t)><i(t)|.
\end{equation}
Differentiating the last expression on time we obtain the equation
\begin{equation}\label{U18}
\frac{d\rho}{dt}=\sum_{i}
\frac{d\omega_{i}[\beta(t)]}{dt}|i(t)><i(t)|-i[H,\rho(\beta)]=d[ln\omega(\beta)]\rho
(\beta)-i[H,\rho(\beta)].
\end{equation}
Where $ln[\omega(\beta)]$ is a row-matrix and $\rho(\beta)$ is a
column-matrix. In such a way we have obtained a prototype of the
 Liouville's equation.

Let's consider some particular cases of importance.
\begin{enumerate}
\item First we consider the process of time
evolution at low energies, or in other words, when $\beta(0)
\approx 0$, $\beta(t)\approx 0$ and $t \to \infty$. Then it is
clear that $\omega_{i}(\beta)\approx \omega_{i} \approx constant$.
The first term in (\ref{U18}) vanishes and we obtain the
Liouville's equation.
\item We obtain also the Liouville's equation when we turn from
inflationary  to large-scale. Within the inflationary approach the
scale $a \approx e^{Ht}$, where $H$ is the Hubble's constant and
$t$ is time. Therefore $\beta \sim e^{-2Ht}$ and when $t$ is quite
big $\beta \to 0$. In other words, $\omega_{i}(\beta) \to
\omega_{i}$, the first term in (\ref{U18}) vanishes and we obtain
again the Liouville's equation.
\item At very early inflationary-process stage or even before it
takes place $\omega_{i}(\beta)$ was not a constant and therefore,
the first term in (\ref{U18}) should be taking into account. This
way we obtain a deviation from the Liouville's equation.
\item At last, let us consider the case when $\beta(0) \approx 0$,
$\beta(t)>0$ where $t \to \infty$. In this case we are going from
low-energy scale to high one and $\beta(t)$ grows when $t \to
\infty$. In this case the first term in (\ref{U18}) is significant
and we obtain an addition to the Liouville's equation in the form
$$d[ln\omega(\beta)]\rho(\beta).$$ This case could take place when
matter go into a black hole and is moving in direction of the
singularity (to the Planck's scale).
\end{enumerate}

 \section{Singularity, Entropy and Information Loss in Black
Holes} It follows to note that remark II in section 3 about
complex meaning assumed by density pro-matrix at the point with
fundamental length has straightforward relation with the
singularity problem and cosmic censorship in General Theory of
Relativity \cite{r11}. Indeed, considering, for instance, a
Schwarzchild's black hole (\cite{r12}) with metrics:
\begin{equation}\label{U19}
 ds^2 = - (1 - \frac{2M}{r}) dt^2 +
\frac{dr^2}{(1 - \frac{2M}{r})} + r^2 d \Omega_{II}^2,
\end{equation}
we obtain, as it is well-known a singularity at the point $r=0$.
In our approach this corresponds to the point with fundamental
length ($r=l_{min}$). At this point we are not able to measure
anything, since at this point $\beta=1$ and $Sp[\rho (\beta)]$
becomes complex. Thus, we carry out a measurement, starting from
the point $r=2l_{min}$ corresponding to $\beta=1/4$. Consequently,
the initial singularity $r=l_{min}$, which cannot be measured, is
hidden of observation. This could confirm the hypothesis of cosmic
censorship in this concrete case. This hypothesis claims that "a
naked singularity cannot be observed". Thus, QMFL in our approach
feels the singularity. (In comparison with QM, which does not feel
it).

Statistical entropy, connected with density pro-matrix and
introduced in the remark V, section 3
$$S_{\beta}=-Sp[\rho(\beta)\ln(\rho(\beta))],$$  may be
interpreted as density of entropy on unity of minimal square
$l^{2}_{min}$ depending on the scale $x$. It could be quite big
nearby the singularity. In other words, when $\beta\rightarrow
1/4$. This does not contradict the second law of Thermodynamics
since the maximal entropy of a determined object in the Universe
is proportional to the square of their surface $A$, measured in
units of minimal square $l^{2}_{min}$ or Planck's square
$L_{p}^2$, as it was shown in some papers (see, for instance
\cite{r13}). Therefore, in the expanded Universe since surface $A$
grows, then entropy does not decrease.

Obtained results enable one to consider anew the information loss
problem associated with black holes \cite{r14,r15}, at least, for
the case of "mini" black holes. Indeed, when we consider these
black holes, Planck's scale is a factor. It was shown that entropy
of matter absorbed by a black hole at this scale is not equal to
zero, supporting the data of R.Myers \cite{r16}. According to his
results, a pure state cannot form a black hole. Then,  it is
necessary to reformulate the problem per se, since so far in all
publications on information paradox zero entropy at the initial
state has been compared to  nonzero entropy at the final state.
According to our analysis at the Planck's scale there is not
initial zero entropy and "mini" black holes with masses of the
order $M_{pl}$ should not radiate at all. Similar results using
another approach were deduced by A.D.Hefler\cite{r30}: "p.1...The
possibility that non-radiating "mini" black holes should be taken
seriously; such holes could be part of the dark matter in the
Universe". It follows to note, that in \cite{r30}, the main
argument in favor of the existence of non-radiating "mini" black
holes is founded under the consideration of quantum gravity
effects. In our analysis these effects are considered implicitly
since any approach in quantum gravity leads to the
fundamental-length concept \cite{r26}. Besides, it should be noted
that in some recent papers for all types of black holes QMFL with
GUR is considered at the very beginning \cite{r17},\cite{r31}. As
a consequence of this approach, stable remnants with masses of the
order of Planck's ones $M_{pl}$ emerge during the process of black
hole evaporation. From here it follows that black holes should not
evaporate fully. We arrive to the conclusion that results given in
\cite{r12, r18} are correct only in the semi-classical
approximation and they should not be applicable to the quantum
back hole analysis. Based on our results, we can elucidate (at
least qualitatively) the problem associated with information loss
on black holes formed when a star collapses. Actually, near the
horizon of events the entropy of an approximately pure state is
practically equal to zero: $S^{in}=-Sp[\rho \ln(\rho)]$ that is
associated with the value $\beta \mapsto 0$. When approaching the
singularity $\beta>0$ (i.e. on Plank's scale), its entropy is
nonzero for $S_{\beta}=-Sp[\rho (\beta)\ln(\rho(\beta))]$.
Therefore, in such a black hole the entropy increases, whereas
information is lost.

On the other hand, from the results obtained above, at least at
the qualitative level, it can be clear up the answer to the
question how may be information lost in big black holes, which are
formed as result of star collapse. Our point of view is closed to
the R. Penrose's one \cite{r19}. He considers  that information in
black holes is lost when matter meets a singularity. In our
approach information loss takes place in the same form. Indeed,
near to the horizon of events an approximately pure state with
practically equal to zero initial entropy
$S^{in}=-Sp[\rho\ln(\rho)]$, which corresponds to $\beta \to 0$,
when approaching a singularity (in other words, is reaching the
Planck's scale) gives yet non zero entropy
$S_{\beta}=-Sp[\rho(\beta)\ln(\rho(\beta))]>0$ when $\beta >0$.
Therefore, entropy increases and information is lost in this black
hole. We can (at the moment, also at the qualitative level)
evaluate entropy of black holes. Indeed, starting from density
matrix for a pure state at the "entry" of a black hole
$\rho_{in}=\rho_{pure}$ with zero entropy $S^{in}=0$, we obtain,
doing a straightforward "naive" calculation (this means that
(\ref{U13}) is considered as an exact relation). Then, at the
singularity in the black hole density pro-matrix
$Sp[\rho_{out}]=1/2)$ for $\beta=1/4$ with entropy
$$S^{out}=S_{1/4}=-1/2 \ln1/2 \approx 0.34657.$$
 Taking into account that total entropy of a
black hole is proportional to quantum area of surface A, measured
in Planck's units of area $L_{p}^2$ \cite{r20}, we obtain the
following value for quantum entropy of a black hole:
\begin{equation}\label{U20}
S_{BH}= 0.34657 \frac{A}{L_{p}^2}
\end{equation}

This formula differs from the well-known one given by
Bekenstein-Hawking for black hole entropy $S_{BH}=\frac{1}{4}
\frac{A}{L_{p}^2}$ \cite{r21}. This result was obtained in the
semi-classical approximation. At the present moment quantum
corrections to this formula are object of investigation
\cite{r22}. As it was yet above-mentioned we carry out a
straightforward calculation. Otherwise, using the ansatz of the
remark III in section 3 and assuming that spur of density
pro-matrix is equal to $Sp[\rho^{*}(\beta)]=exp(-\beta)$, we
obtain for $\beta=1/4$ that entropy is equal to $$S^{*
out}=S^{*}_{1/4}=-Sp[exp(-1/4)\ln exp(-1/4)]\approx 0.1947,$$ and
consequently we arrive to the value for entropy
\begin{equation}\label{U21}
S_{BH} = 0.1947 \frac{A}{L_{p}^2}
\end{equation}
that is nearest to the result obtained in \cite{r22}. Our
approach, leading to formula (\ref{U21}) is at the very beginning
based on the quantum nature of black holes. Let us to note here
that in the approaches, used up to now to modify Liouville's
equation, due to information paradox \cite{r23}, the added member
appearing in the right side of (\ref{U18}) has the form
$$-\frac{1}{2}\sum_{\alpha \beta \neq 0}
(Q^{\beta}Q^{\alpha}\rho+\rho Q^{\beta}Q^{\alpha}-2 Q^{\alpha}\rho
Q^{\beta}),$$ where $Q^{\alpha}$
  is a full orthogonal set
of Hermitian matrices with $Q^{0} =1$. In this case either
locality or conservation of energy-impulse tensor is broken down.
In the offered in this paper approach, the added member in the
deformed Liouville's equation  has a more natural and beautiful
form in our opinion: $$d[ln\omega(\beta)]\rho (\beta).$$ In the
limit $\beta\to 0$ all properties of QM are conserved, the added
member vanishes and we obtain Liouville's equation.

\section{Bekenstein-Hawking Formula}
  Whether can the well-known semiclassical Bekenstein-Hawking formula for Black Hole
  entropy be obtained within the proposed approach? Let us show how
  to do it. At the moment our deduction is founded on an heuristic
  argumentation. However we are sure a more strong argumentation can be provided.
  To obtain black hole quantum entropy we use the formula
  $S_{\beta}=-Sp[\rho(\beta)\ln(\rho(\beta))]=-<\ln(\rho(\beta))>_{\beta}$
    when $\beta$ takes its maximal meaning ($\beta = 1/4$).
   In this case (\ref{U20}) and (\ref{U21}) can be written as
\begin{equation}\label{U22}
S_{BH} = -<\ln(\rho(1/4))>_{1/4} \frac{A}{L_{p}^2},
\end{equation}
for different $\rho(\beta)$ in  (\ref{U20}) and  (\ref{U21}) but
for the same meaning of $\beta$ ($\beta = 1/4$). Semiclassical
approximation works only at large-scales, therefore measuring
procedure is also defined at the large-scales. In other words, all
mean values must be taken when $\beta = 0$. However, for
operators, whose mean values are calculated the dependence on
$\beta$ must be taking into account since according to the
well-known Hawking's paper \cite{r14} operator of superscattering
$\$$ translates $\$:\rho_{in}\mapsto\rho_{out}$, where in the case
considered $\rho_{in}=\rho_{pure}$ and
$\rho_{out}=\rho_{pure}^{*}(\beta)=
exp(-\beta)\rho_{pure}=exp(-1/4)\rho_{pure}$ in correspondence
with the exponential ansatz of point III, section 3. Therefore we
have
\\
$$S^{semiclass}_{\beta}=-<\ln(\rho(\beta))>$$
\\
and formula for semiclassical entropy for a black hole takes the
form
\begin{equation}\label{U23}
S^{semiclass}_{BH} = -<\ln(\rho(1/4))>
\frac{A}{L_{p}^2}=-<ln[exp(-1/4)]\rho_{pure}>\frac{A}{L_{p}^2}
=\frac{1}{4}\frac{A}{L_{p}^2}
\end{equation}
that coincides with the well-known Bekestein-Hawking formula. It
follows to note that the meaning $\beta = 1/4$  in our approach
appears in section 3 non in an artificial way but as the maximal
meaning for which $Sp\rho(\beta)$ still stays real, according to
(\ref{U13}) and (\ref{U14}). Apparently, if considering
corrections of order higher than 1 on $\beta$, then members from
$O(\beta^{2})$ in the formula for $\rho_{out}$ in (\ref{U15}) can
give quantum corrections \cite{r22} for $S^{semiclass}_{BH}$
(\ref{U23}) in our approach.

\section{Some comments on Shr{\"o}dinger's picture}
A procedure allowing to obtain a theory from the transformation of
the precedent one is named "deformation". This is doing, using one
or a few parameters of deformation in such a way, that the
original theory must appear in the limit, when all parameters tend
to some fixed values. The most clear example is QM being a
deformation of Classical Mechanics. The parameter of deformation
in this case is the Planck's constant $\hbar$. When
$\hbar\rightarrow 0$ QM passages to Classical Mechanics. As it was
indicated above in the remark 1 section 3, we are able to obtain
from QMFL two limits: Quantum and Classical Mechanics. The
described here deformation should be understood as "minimal" in
the sense that we have deformed only the probability
$\omega_{i}\rightarrow \omega_{i}(\beta)$, whereas state vectors
have been not deformed. In a most complete consideration we will
be obligated to consider instead $|i><i|$, vectors
$|i(\beta)><i(\beta)|$ and in this case the full picture will be
very complicated. It is easy to understand how Shrodinger's
picture is transformed in QMFL. The prototype of Quantum
Mechanical normed wave function $\psi(q)$ with
$\int|\psi(q)|^{2}dq=1$ in QMFL is $\theta(\beta)\psi(q)$. The
parameter of deformation $\beta$ assumes the value
$0<\beta\leq1/4$. Its properties are
$|\theta(\beta)|^{2}<1$,$\lim\limits_{\beta\rightarrow
0}|\theta(\beta)|^{2}=1$ and the relation
$|\theta(\beta)|^{2}-|\theta(\beta)|^{4}\approx \beta$ takes
place. In such a way the full probability always is less than 1:
$p(\beta)=|\theta(\beta)|^{2}=\int|\theta(\beta)|^{2}|\psi(q)|^{2}dq<1$
and it tends to 1 when  $\beta\rightarrow 0$. In the most general
case of arbitrarily normed state in QMFL
$\psi=\psi(\beta,q)=\sum_{n}a_{n}\theta_{n}(\beta)\psi_{n}(q)$
with $\sum_{n}|a_{n}|^{2}=1$ the full probability is
$p(\beta)=\sum_{n}|a_{n}|^{2}|\theta_{n}(\beta)|^{2}<1$ and
 $\lim\limits_{\beta\rightarrow 0}p(\beta)=1$.

 It is natural that in QMFL Shrodinger's equation is also
deformed. It is replaced by the  equation
\begin{equation}\label{U24}
\frac{\partial\psi(\beta,q)}{\partial t}
=\frac{\partial[\theta(\beta)\psi(q)]}{\partial
t}=\frac{\partial\theta(\beta)}{\partial
t}\psi(q)+\theta(\beta)\frac{\partial\psi(q)}{\partial t},
\end{equation}
where the second term in the right side generates the Shrodinger's
equation since
\begin{equation}\label{U25}
\theta(\beta)\frac{\partial\psi(q)}{\partial
t}=\frac{-i\theta(\beta)}{\hbar}H\psi(q).
\end{equation}

Here $H$ is the Hamiltonian and the first member is added,
similarly to the member appearing in the deformed Loiuville's
equation and  vanishing when $\theta[\beta(t)]\approx const$. In
particular, this takes place in the low energy limit in QM, when
$\beta\rightarrow 0$.  It follows to note that the above-described
theory  is not time-reversal as QM, since the combination
$\theta(\beta)\psi(q)$ breaks down this property in the deformed
Shrodinger's equation. Time-reversal is conserved only in the low
energy limit, when quantum mechanical Shrodinger's equation is
valid.

\section{Conclusion}
It follows to note, that in some well-known papers on GUR and
Quantum Gravity (see for instance \cite{r1,r2,r3,r4,r24}) there is
not any mention about the measuring procedure. However, it is
clear that this question is crucial and it cannot be ignored or
passed over in silence. We would like to remark, that the
measuring rule used in (\cite{r29}, formula (5)) coincide with
ours. Taking into account this state of affairs we propose in this
paper a detailed treatment of the problem of measurement. In the
paper the measuring rule (\ref{U6}) is proposed as a good initial
approximation to the exact measuring procedure in QMFL.
Corrections to this procedure could be defined by an adequate and
fully established description of the space-time foam (see
\cite{r25}) at Planck's scale. On the other hand, as it was noted
in  (see \cite{r26}) all known approaches
 dealing with Quantum Gravity
one way or another lead to the notion of fundamental length.
Involving that notion too, GUR (\ref{U2}) are well described by
the inflation model \cite{r27}. Therefore, it seems impossible to
understand physics on Planck's scale disregarding the notion of
fundamental length. One more aspect of this problem should be
considered. As it was noted in \cite{r28}, advancement of a new
physical theory implies the introduction of a new parameter and
deformation of the precedent theory by this parameter. In essence,
all these deformation parameters are fundamental constants: $G$,
$c$ and $\hbar$ (more exactly in \cite{r28} $1/c$ is used instead
of $c$). As follows from the above results, in the problem from
\cite{r28} one may redetermine, whether a theory we are seeking is
the theory with fundamental length involving these three
parameters by definition: $L_{p}=\sqrt\frac{G\hbar}{c^3}$. Notice
also that the deformation introduced in this paper is stable in
the sense indicated in \cite{r28}.

\noindent The present paper is  the continuation and correction of
the  \cite{r33}.


\end{document}